# Tunable interfacial Rashba spin-orbit coupling in asymmetric $Al_xIn_{1-x}Sb$/InSb/CdTe quantum well heterostructures

Hanzhi Ruan[1,#], Zhenghang Zhi[1,3,4,#], Yuyang Wu[2,#], Jiuming Liu[1], Puyang Huang[1], Shan Yao[1], Xinqi Liu[5,6], Chenjia Tang[5,6], Qi Yao[5,6], Lu Sun[1], Yifan Zhang[1], Yujie Xiao[5], Renchao Che[2,*] and Xufeng Kou[1,6,*]

[1] *School of Information Science and Technology, ShanghaiTech University, Shanghai, 201210, China*

[2] *Department of Materials Science, Fudan University, Shanghai, 20043, China*

[3] *Shanghai Institute of Microsystem and Information Technology, Chinese Academy of Sciences, Shanghai 200050, China*

[4] *University of Chinese Academy of Science, Beijing 101408, China*

[5] *School of Physical Science and Technology, ShanghaiTech University, Shanghai, 201210, China*

[6] *ShanghaiTech Laboratory for Topological Physics, ShanghaiTech University, Shanghai 200031, China*

[#] These authors contributed equally to this work

*e-mail: kouxf@shanghaitech.edu.cn; rcche@fudan.edu.cn

**Abstract**

We report the manipulation of the Rashba-type spin-orbit coupling (SOC) in molecular beam epitaxy-grown $Al_xIn_{1-x}Sb/InSb/CdTe$ quantum well heterostructures. The effective band bending warrants a robust two-dimensional quantum confinement effect, and the unidirectional built-in electric field arisen from the asymmetric hetero-interfaces leads to a pronounced Rashba SOC strength. By tuning the Al concentration in the top $Al_xIn_{1-x}Sb$ barrier layer, the optimal structure of $x = 0.15$ exhibits the largest Rashba coefficient of 0.23 eV·Å as well as the highest low-temperature electron mobility of 4400 $cm^2·V^{-1}·s^{-1}$. Moreover, quantitative investigations of the weak anti-localization effect further justify the prevailing D'yakonov-Perel (DP) spin relaxation mechanism during the charge-to-spin conversion process. Our results underscore the importance of quantum well engineering in shaping the magneto-resistance responses, and the narrow bandgap semiconductor-based heterostructures may serve as a reliable framework for designing energy-efficient spintronic applications.

The spin-orbit coupling (SOC) is a fundamental concept in quantum mechanics that plays an essential role in tailoring the electron spin state and constructing energy-efficient spintronic applications.[1-6] In bulk materials with large SOC (*e.g.*, heavy metals, narrow bandgap semiconductors, and topological quantum materials), the charge current can be converted to spin current through the spin Hall effect, henceforth enabling the spin generation and detection by electric field.[7-11] Alternatively, another prominent manifestation of SOC is the interfacial Rashba effect, which occurs in heterogeneous systems with a natural broken inversion symmetry (*e.g.*, metal/semiconductor/oxide-based two-dimensional quantum well structures).[12-15] In such heterostructures, the interfacial SOC is closely associated with the potential gradient (*i.e.*, the built-in electric field) induced by the band bending, and its strength can be precisely modulated through the gate bias without necessitating an external magnetic field. Accordingly, the key to design energy-efficient Rashba-type spintronic devices relies on the dedicated design of heterostructures with pronounced band confinement and high carrier mobility.[16]

Among all material candidates, narrow bandgap III-V compound semiconductors such as InSb and InAs are well known for their light effective mass, large *g*-factor, high carrier mobility, and inherently strong SOC. Moreover, in our previous study, we have reported the integration of InSb and CdTe (*i.e.*, II-VI semiconductor with the energy bandgap of ~1.5 eV[17]) not only warrants a high-quality heterostructure with negligible crystallographic defects owing to their identical lattice constants, but also hosts a well-defined hetero-interface with a large built-in electric field[18]. In this context, the resulting large Rashba-type SOC is highly advantageous for efficient spin manipulation.[19] For instance, evident non-reciprocal charge transport and spin-orbit torque-driven magnetization switching have been demonstrated in InSb/CdTe-based protype devices up to room temperature.[18, 20-22]

Inspired by the aforementioned progress, in this study, we report the introduction of the $Al_xIn_{1-x}Sb$ barrier layer to form the asymmetric $Al_xIn_{1-x}Sb$/InSb/CdTe quantum well (QW) structure which manipulate the charge and spin transports. Benefiting from an improved quantum confinement, this sandwich structure exhibits both a boosted electron mobility and a pronounced weak anti-localization (WAL) effect from the temperature-dependent magneto-transport responses. Moreover, by varying the Al content of the $Al_xIn_{1-x}Sb$ barrier layer, the overall interfacial Rashba SOC strength and spin relaxation length are effectively modulated. Our results highlight the importance of QW structural engineering in optimizing the electrical and spin performance, and validate the narrow bandgap semiconductor-based heterostructures as a suitable platform for advancing energy-efficient and low-power spintronic applications.

Experimentally, the $Al_xIn_{1-x}Sb$/InSb/CdTe trilayer quantum well heterostructures were grown on the 3-inch GaAs(111)B substrates by molecular beam epitaxy (MBE). As elaborated in our previous report,[23] a 400 nm insulating CdTe thin film was firstly adopted as the buffer layer to quickly release the stress from the GaAs substrate as well as to provide the lattice-matching condition for the subsequent InSb (30 nm) epitaxial growth. In order to form the quantum well structure, another 40 nm $Al_xIn_{1-x}Sb$ layer was introduced as the top barrier, where the Al concentration was carefully chosen from 0 to 18%, and sharp 2D streaky reflection high energy electron diffraction (RHEED) patterns are observed during the entire $Al_xIn_{1-x}Sb$/InSb layer growth, as illustrated in Fig. 1(b). Finally, the 2nm highly-doped InSb passivation layer was designed to prevent the sample from oxidation while helping reduce the contact resistance. After sample growth, high-resolution scanning transmission electron microscope (HR-STEM) was carried out to visualize the cross-sectional crystalline structure of the $Al_xIn_{1-x}Sb$/InSb/CdTe samples with a well-ordered atom configuration, sharp interfaces, and the absence of macroscopic thread dislocations, as shown in Figs. 1(c)-(e). Furthermore, the built-in electric field $E_{bi}$ was examined by the differential phase contrast (DPC) electron microscopy, which confirms that the directions of the built-in electric fields at both hetero-interfaces point to the $-z$-axis (purple arrows of Fig. 1(f),).

In addition to material characterizations, the band diagram of the $Al_xIn_{1-x}Sb$/InSb/CdTe QW was quantified by TCAD simulation which is based on the Poisson-Schrödinger compact model. The right panel of Fig. 1(a) exemplifies the band edge profiles of under two Al-content conditions ($x = 0.1$ and $x = 0.15$). It is seen that the introduction of Al gives rise to a conduction band offset of $\Delta E_C = 0.114$ eV at the $Al_{0.1}In_{0.9}Sb$/InSb interface, and the formation of the quantum well guarantees an effective confinement of the two-dimensional electron transport. As the Al content further increases from 10% to 15%, the enlarged $\Delta E_C$ causes the Fermi level $E_F$ to move upward in the quantum well, which in turn leads to a higher carrier concentration.[24, 25] More importantly, in agreement with the DPC data of Fig. 1(d), our TCAD simulation results also suggest the presence of a depletion-type heterojunction at the InSb/CdTe interface, while the band bending introduces an inversion layer at the $Al_xIn_{1-x}Sb$/InSb interface at equilibrium. Under such circumstances, the corresponding built-in potential gradients have the same direction at the two hetero-interfaces (*i.e.*, blue arrows in the right panel of Fig. 1(a)), thereby enabling the same spin chirality within the $Al_xIn_{1-x}Sb$/InSb/CdTe heterostructures.

To investigate the electrical properties of the asymmetric $Al_xIn_{1-x}Sb$/InSb/CdTe structure, the MBE-grown 3-inch samples were fabricated into the μm-size top-gated six-terminal Hall bar devices (*i.e.*, with typical channel geometry of $L = 40$ μm and $W = 20$ μm) by standard nano-fabrication process.[18] Afterwards, temperature-dependent magneto-transport measurements were conducted on a set of $Al_xIn_{1-x}Sb$(40nm)/InSb(30 nm)/CdTe(400 nm) samples with $x = 0.1$ (Sample A), 0.13 (Sample B), 0.15 (Sample C) and 0.18 (Sample D). From the summarized Table I, it is found that both the carrier mobility and density disclose a non-monotonic correlation with the Al content in the top $Al_xIn_{1-x}Sb$ barrier layer. Consistent with other reports,[26-28] as $x$ increases from 0.1, the better electron confinement helps to promote the low-temperature electron mobility which reaches the peak value of $\mu = 4400$ cm$^{-2}\cdot$V$^{-1}\cdot$s$^{-1}$ in the $Al_{0.15}In_{0.85}Sb$/InSb/CdTe sample. On the other hand, a higher Al-to-In ratio ($x = 0.18$) results in a smaller lattice constant compared to InSb, which may bring about more interfacial defects and alloy scatterings to degrade the overall channel mobility at low temperatures.[28-30]

Governed by the interplay between the electron spin propagation and the Berry phase, the SOC strength can be characterized by the weak-antilocalization (WAL) effect.[31-33] In particular, the Rashba effect-induced spin-momentum locking mechanism ensures a destructive quantum interference as the interfacial electrons move around self-intersecting scattering paths (*i.e.*, reduced backscattering probability), therefore leading to an enhanced electron conductivity at zero magnetic field; yet the WAL response would fall off quickly in the presence of a perpendicular magnetic field owning to the breaking of time-reversal symmetry.[34] Accordingly, Fig. 2(a) displays the normalized longitudinal magnetoconductance (MC) curves $\Delta\sigma_{xx} = \sigma_{xx}(B) - \sigma_{xx}(0)$ of Samples A-D at $T = 1.5$ K, where the emergence of four positive MC cusps in the low-field region (–0.2 T ≤ $B$ ≤ 0.2 T) manifest the robust WAL phenomena in the $Al_xIn_{1-x}Sb/InSb/CdTe$ QW system. Strikingly, compared to other three samples, Sample C ($x = 0.15$) exhibits the most pronounced WAL-related conductance correction contribution, characterized by the largest MC cusp amplitude of $\Delta\sigma_{xx} = 0.27e^2/h$ (*i.e.*, indicative of a long spin coherent length) and widest magnetic-filed range of 120 mT (*i.e.*, which suggests an optimized overall SOC strength).

To quantify the Rashba coefficient $\alpha_R$ and phase-coherent length $l_\varphi$ in the $Al_xIn_{1-x}Sb/InSb/CdTe$ QW samples, we utilized the Iordanskiii-Lyanda-Geller-Pikus (ILP) model based on the D'yakonov-Perel (DP) spin relaxation mechanism.[33, 35-37] Specifically, the low-field MC curve is expressed as:

$$\Delta\sigma_{xx}(B) = \frac{-e^2}{4\pi^2\hbar}\left\{3C + \frac{1}{a_0} + \frac{2a_0 + 1 + \frac{B_{so}}{B}}{a_1\left(a_0 + \frac{B_{so}}{B}\right) - 2\frac{B_{so}}{B}}\right.$$
$$- \left(\sum_{n=1}^{\infty}\frac{3}{n} - \frac{2a_n^2 + 2\frac{B_{so}}{B}a_n - 1 - 2(2n+1)\frac{B_{so}}{B}}{\left(a_n + \frac{B_{so}}{B}\right)a_{n-1}a_{n+1} - 2\frac{B_{so}}{B}[(2n+1)a_n - 1]}\right) + 2ln\left(\frac{B_{tr}}{B}\right)$$
$$\left. + \Psi\left(\frac{1}{2} + \frac{B_\phi}{B}\right)\right\}$$

where $e$ is the electron charge, $\hbar$ is the reduced Planck constant, $C$ is the Euler's constant, $\Psi$ represents the digamma function, the parameter $a_n = n + \frac{1}{2} + \frac{B_\varphi}{B} + \frac{B_{so}}{B}$, the $B_{so}$, $B_\varphi$, $B_{tr}$ denote the characteristic fields for the Rashba interaction, phase-coherent transport, and momentum scattering process in the system. As exemplified in Fig. 2(b), the ILP model (red dotted line) manages to capture the MC line-shape of the $x = 0.15$ sample in the entire $[-0.2T, +0.2 T]$ region. According to the DP mechanism,[38, 39] the spin relaxation of the conducting electron in the high-mobility system should predominantly arise from the interplay between the Rashba effect-induced pseudo-magnetic field $B_{SO}$ and the carrier scatterings, and its spin precession is driven by the Larmor frequency vector $\boldsymbol{\Omega}(\boldsymbol{k}_\parallel) = \frac{1}{\hbar}[\alpha_R k_y - \gamma k_x^2 k_y), (\alpha_R k_x - \beta k_y + \gamma k_x^2 k_y), 0]$ in the $x$-$y$ plane (i.e., where $\beta$ and $\gamma$ are the coefficients related to the Dresselhaus effect).[40, 41] As a result, the spin precession frequency is positively correlated with the Rashba SOC strength, which means more momentum scattering events (i.e., the change of the wave vector $\boldsymbol{k}_\parallel$) causes slower spin relaxation rates. From the measured gate-dependent MC data of our $Al_{0.15}In_{0.85}Sb/InSb/CdTe$ sample at $T = 1.5$ K, the spin-orbit relaxation time $\tau_{so} = \hbar/4eDB_{so}$ ($D$ is the diffusion constant) is found to be inversely proportional to the momentum scattering time $\tau_{tr} = \mu m^*/e$ (Fig. 2(c)), hence validating the dominant DP spin relaxation mechanism at cryogenic temperatures. As a comparison, we also tried to apply the Hikami-Larkin-Nagoka (HLN) model which is based on the Elliot-Yafet (EY) mechanism, yet failed to achieve a good fitting to the experimental result (blue dashed line of Fig. 2(b)). Besides, Fig. 2(d) displays the zero-field conductivity $\sigma_{xx}$ as a function of temperature, and the linear $\sigma_{xx} - \log(T)$ curves in the low-temperature region ($T < 10$ K) unveil the two-dimensional transport characteristics of the $Al_xIn_{1-x}Sb/InSb/CdTe$ quantum wells under the influence of the quantum interference effects.[42-44]

Based on the WAL fitting results, we were able to quantify the Rashba SOC strength in our $Al_xIn_{1-x}Sb/InSb/CdTe$ heterostructures. In fact, the extracted effective fields $B_{so}$ and $B_\varphi$ from the ILP model can be converted to the Rashba coefficient $\alpha_R = \sqrt{e\hbar^3 B_{so}}/m^*$ ($m^*$ is the electron effective mass)[45, 46] and phase-coherent length $l_\varphi = \sqrt{\hbar/4eB_\varphi}$, respectively.[18,47] In this regard, Figs. 3(a)-(b) display the temperature-dependent $\alpha_R$ and $l_\varphi$ values of Samples A-D, where several unique features are identified.

First, regardless the Al content in the top $Al_xIn_{1-x}Sb$ barrier, all the $\alpha_R$-$T$ curves exhibit the same evolution trend: the calculated $\alpha_R$ values remain almost constant in the high-temperature region (20 K < $T$ < 80 K), whereas they all experience a dramatic increase when $T$ < 20 K. Considering that the bulk conduction (*i.e.*, through the $Al_xIn_{1-x}Sb$ and InSb channels) is suppressed at deep cryogenic temperatures, the enhanced interfacial charge transport would contribute to a larger overall Rashba coefficient in our QW samples (*i.e.*, the degrees of change in carrier density and $\alpha_R$ are comparable to each other, as shown in the inset of Fig. 3(a)). Secondly, the phase-coherent length $l_\varphi$ monotonically decreases with temperature, yet its scaling behavior varies in samples with different Al% values. As depicted in Fig. 3(b), Samples A-C generally follow the same $l_\varphi \propto T^{-\frac{p}{2}}$ scaling law with the exponent $p = 1$ as long as $T > 15$ K, which imply that the electron-electron interaction is the predominant scattering mechanism in our $Al_xIn_{1-x}Sb$/InSb/CdTe heterostructures.[48,49] On the other hand, except for sample C ($x = 0.15$), the phase-coherent lengths in all the other samples gradually saturate towards constant values ($l_\varphi^{sat}$) when $T < 5$ K. Here, we need to point out that such a saturation behavior has also been observed in other heterostructures (*e.g.*, Si/SiGe,[0, 51] GaAs/AlGaAs,[52] GaAs/InGaAs[53]) and disordered/polycrystalline metals, probably due to the saturated diffusion constant (*i.e.*, carrier mobility) in the diffusive transport regime at ultra-low temperatures (*i.e.*, $l_\varphi^{sat} \propto D^\alpha$, where $\alpha \geq 1$). Intriguingly, it is also found that although the measured overall carrier mobility of Sample D is comparable to that of Samples A-B, yet its saturated $l_\varphi^{sat}$ value is only 45% of the lightly-doped counterparts. Given that the phase-coherent length is intrinsically correlated with the spin-polarized interfacial carriers, the reduction of $l_\varphi^{sat}$ in Sample D thereafter confirms the presence of lattice mismatch-induced impurity scattering and additional defect-related dephasing at the $Al_{0.18}In_{0.82}Sb$/InSb interface. Furthermore, Figs. 3(c)-(d) offer clear visualizations that both $\alpha_R$ and $l_\varphi$ simultaneously reach the optimal states in the $x = 0.15$ case (*i.e.*, corresponding to $\alpha_R = 0.23$ eV · Å and $l_\varphi = 680$ nm) which makes the best trade-off between the quantum well confinement, built-in electric field, and interfacial scattering in our QW system.

Finally, the comparisons of the overall Rashba coefficients in various strongly spin-orbit coupled material systems are presented in Fig. 4. Thanks to a small electron effective mass, large band offset $\Delta E_C$, and strong effective magnetic field empowered by the large band-bending at the hetero-interface, our $Al_{0.15}In_{0.85}Sb/InSb/CdTe$ sample shows a much larger $\alpha_R$ compared to other semiconductor heterostructures (InAs/GaSb[54] and InAs/AlGaSb[55]), oxide 2DEGs ($LaAlO_3/SrTiO_3$[56]), and 2D materials (Te[48] and $MoS_2$[57]). Concurrently, it is noted that the extracted Rashba coefficient in our system is 50% higher than the AlInSb/InSb/AlInSb counterpart.[58] This is because in symmetric QW structures, the opposite directions of the built-in electric fields at the top and bottom interfaces would give rise to two spin-polarized conduction channels with clockwise and counter-clockwise chirality, which inevitably lower the overall Rashba SOC strength. In contrast, as revealed in Fig. 1, the asymmetric AlInSb/InSb and InSb/CdTe configurations ensure the same $B_{SO}$ direction, therefore resulting in an enhanced Rashba coefficient.

In conclusion, we have demonstrated the use of the $Al_xIn_{1-x}Sb/InSb/CdTe$ quantum well heterostructures to optimize both the carrier mobility and interfacial Rashba-type SOC strength. The $Al_{0.15}In_{0.85}Sb/InSb/CdTe$ film stack exhibited the highest Rashba coefficient of 0.23 eV·Å, spin relaxation length of 680 nm, and electron mobility of 4400 $cm^2 \cdot V^{-1} \cdot s^{-1}$ at $T = 1.5$ K, all of which manifest the advantages of asymmetric narrow bandgap semiconductor-based QW systems for charge-to-spin conversion. By applying modulation doping and/or remote $\delta$-doping strategies in the top $Al_xIn_{1-x}Sb$ layer, we may be able to confine more electrons at the hetero-interface (*i.e.*, which may enhance the overall $\alpha_R$) and to significantly boost the carrier mobility. Additionally, if such a high-quality $Al_xIn_{1-x}Sb/InSb/CdTe$ wafer is further integrated with another magnetic layer, the large spin-orbit torque (*i.e.*, which is proportional to $\alpha_R$) inserted from the spin-generation channel would drive the magnetization switching with an ultra-low threshold current, therefore facilitating the development of low-power, high-speed magnetic memory applications.

**Acknowledgements**

This work is supported by the National Key R&D Program of China (Contract No. 2021YFA0715503), the National Natural Science Foundation of China (Grant No. 92164104), Zhangjiang Lab Strategic Program, the Major Project of Shanghai Municipal Science and Technology (Grant No. 2018SHZDZX02), and the ShanghaiTech Material Device and Soft Matter Nano-fabrication Labs (SMN180827).

**Data Availability**

AIP Publishing believes that all datasets underlying the conclusions of the paper should be available to readers. We encourage authors to deposit their datasets in publicly available repositories (where available and appropriate) or present them in the main manuscript. All research articles must include a data availability statement informing where the data can be found. By data we mean the minimal dataset that would be necessary to interpret, replicate and build upon the findings reported in the article. The data that support the findings of this study are available from the corresponding author upon reasonable request.

**Figure Caption:**

FIG. 1. (a) Left: Schematic of the InSb(2 nm)/Al$_x$In$_{1-x}$Sb(40 nm)/InSb(30 nm)/CdTe(400 nm)/GaAs heterostructures. The patterned six-terminal hall bar device is etched till the insulating CdTe buffer layer. Right: The band diagram of the asymmetric Al$_x$In$_{1-x}$Sb/InSb/CdTe quantum well structure by TCAD simulation. (b) *In-situ* RHEED patterns of the as-grown surface after the deposition of the 30nm InSb channel layer and 40nm Al$_x$In$_{1-x}$Sb ($0 < x \leq 0.18$) barrier layer. (c) High-angle annular dark field (HADDF) image shows the well-ordered crystalline configuration. Both the AlInSb/InSb and InSb/CdTe interfaces are distinguished without notable defects. (d-e) The magnified HR-STEM images of the AlInSb/InSb and InSb/CdTe interfaces in reference to the marked white boxes in (c). (f) Differential phase contrast mapping for the Al$_{0.15}$In$_{0.85}$Sb/InSb/CdTe sample confirms that the built-in electric fields at both hetero-interfaces have the same direction. The color-wheel designates the direction of the electric filed.

FIG. 2. (a) Magneto-conductance results of four Al$_x$In$_{1-x}$Sb/InSb/CdTe samples where pronounced WAL cusps are observed in the low magnetic field region at $T = 1.5$ K. (b) Fitting the measured WAL data of Sample C with ILP (red dotted line) and HLN (blue dashed line) models. (c) Spin-orbit relaxation time $\tau_{SO}$ as a function of the momentum-scattering rate $1/\tau_{tr}$ for Sample C. Inset: Illustration of the top-gated hall bar device. (d) Temperature-dependent longitudinal conductance of Sample A, Sample C and Sample D. The linear $\sigma_{xx} - \log(T)$ scaling law indicates the two-dimensional transport feature at cryogenic temperatures.

FIG. 3. Temperature-dependent (a) Rashba coefficient $\alpha_R$ and (b) phase coherent length $l_\varphi$ for the Al$_x$In$_{1-x}$Sb/InSb/CdTe samples. Both parameters are extracted from the measured magneto-conductance curves, and the black dashed line in Fig. 3(b) represents the ideal dephasing scaling law of $l_\varphi = T^{-0.5}$ governed by the electron-electron interaction. Inset of Fig. 3(a): The evolution trends of the normalized carrier density and Rashba coefficient of Sample C ($x = 0.15$) with respect to the temperature variation.

Comparisons of (c) Rashba coefficient and (d) phase coherent length values among Samples A-D at $T =$ 1.5 K (red circles) and 20 K (blue squares).

FIG. 4. Comparisons of the Rashba coefficients in various strong SOC material systems.

TABLE I. Carrier mobility and density data in $Al_xIn_{1-x}Sb$(40 nm)/InSb(30 nm)/CdTe(400 nm) samples at 1.5 K. The Al contents are $x =$ 0.1 (Sample A), 0.13 (Sample B), 0.15 (Sample C), and 0.18 (Sample D), respectively.

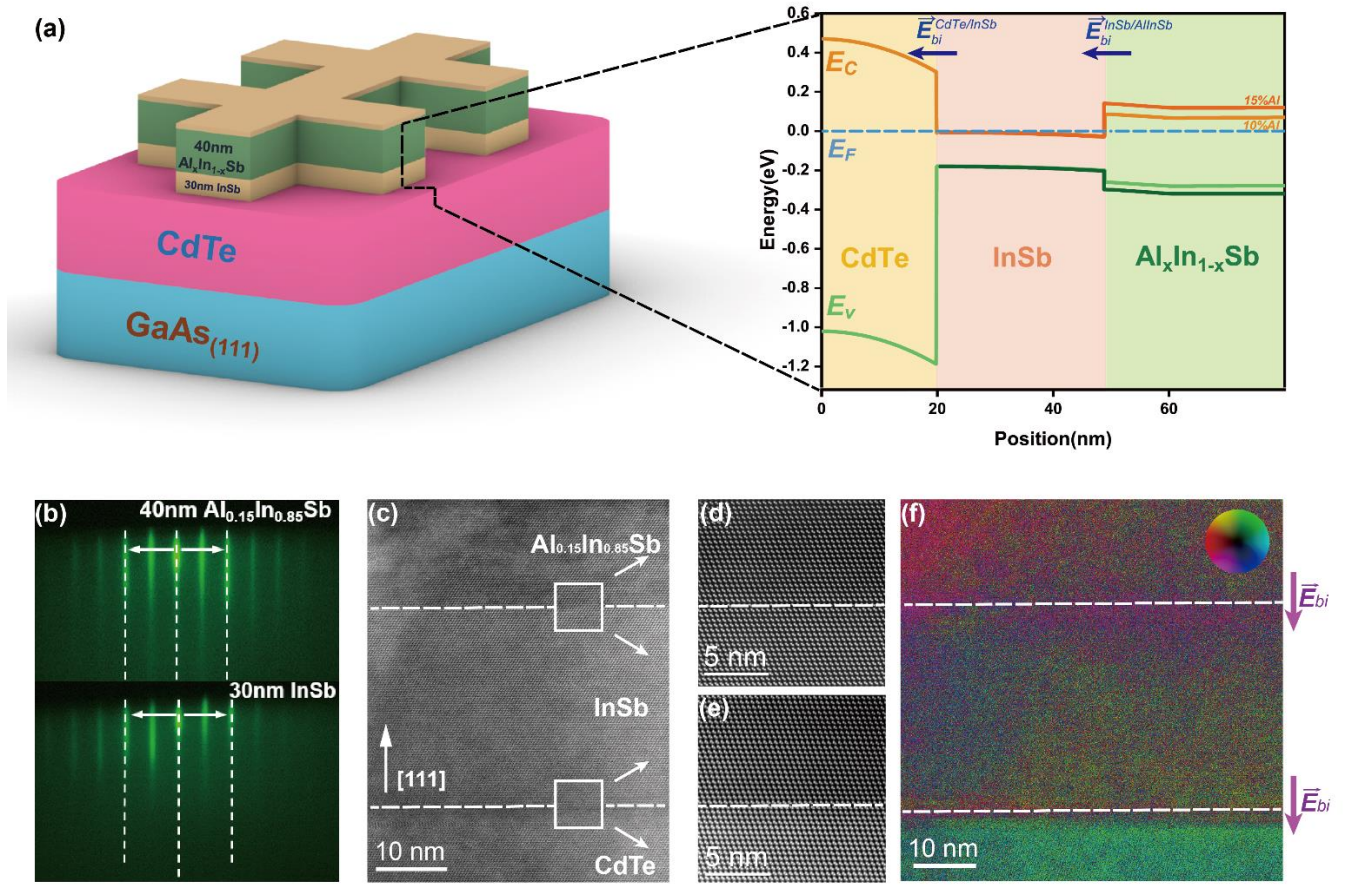

FIG. 1. Zhi *et al.*

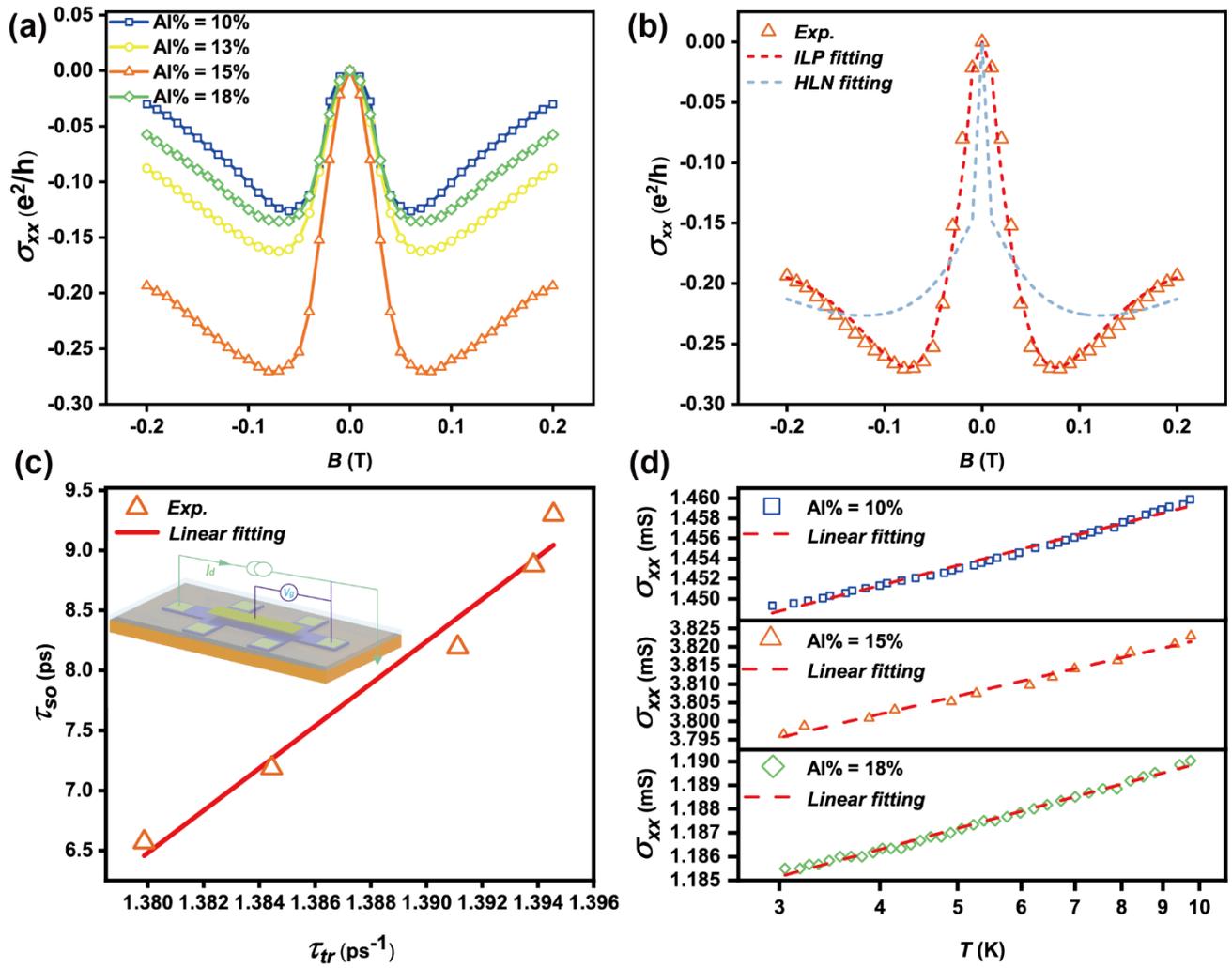

FIG. 2. Zhi *et al.*

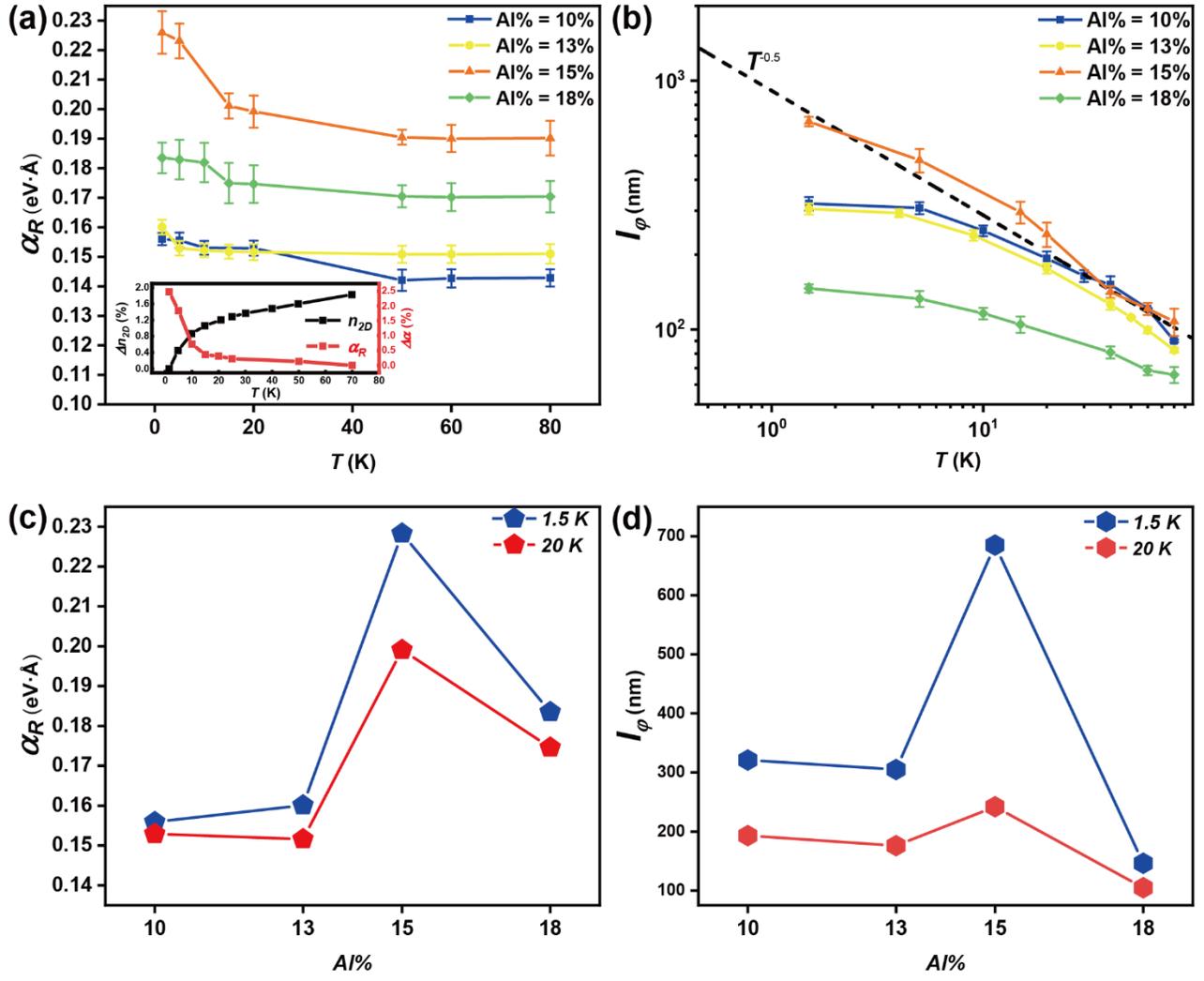

FIG. 3. Zhi *et al.*

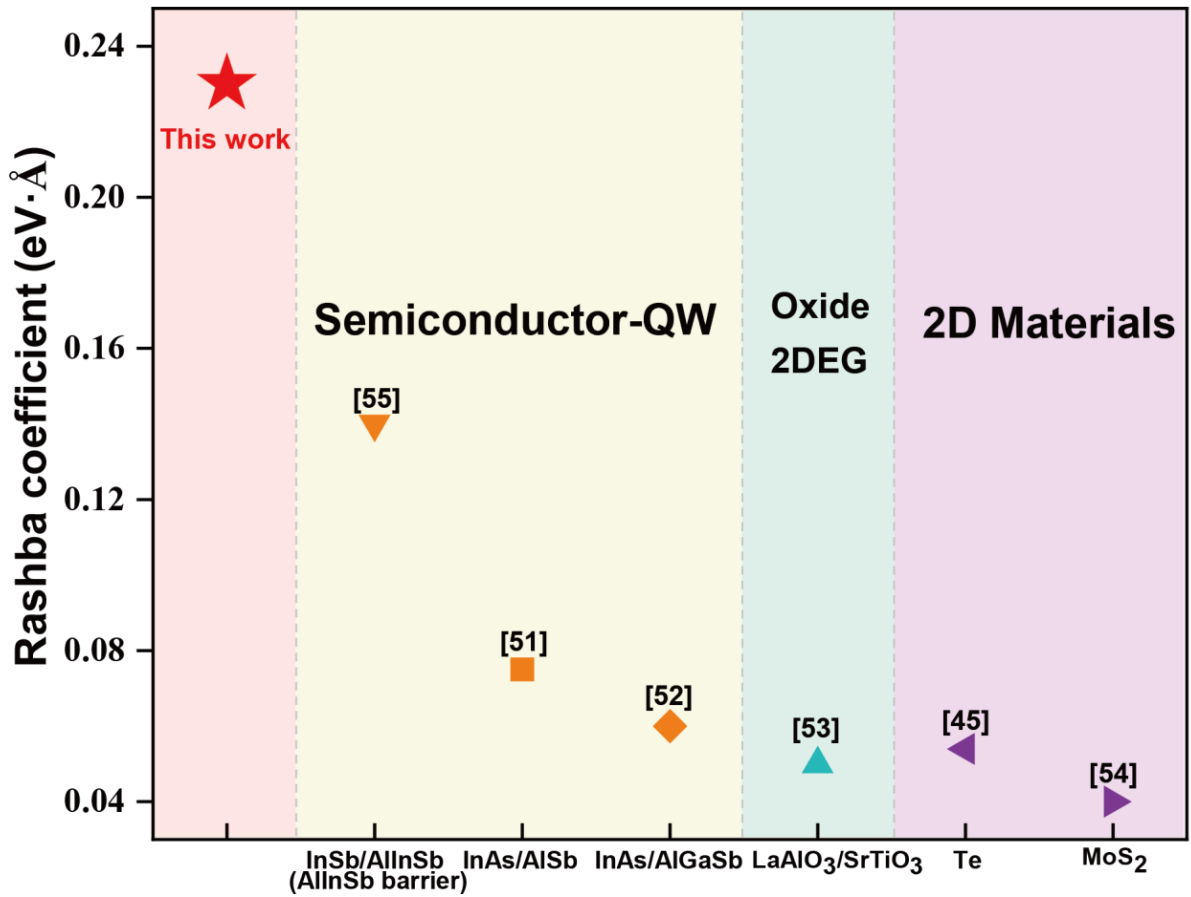

FIG. 4. Zhi *et al.*

|  | Al concentration (%) | $\mu$ at 1.5 K (cm$^2$/Vs) | $n$ at 1.5 K (cm$^{-2}$) |
|---|---|---|---|
| *Sample A* | 10 | 3440 | $4 \times 10^{12}$ |
| *Sample B* | 13 | 3650 | $5.07 \times 10^{12}$ |
| *Sample C* | 15 | 4400 | $5.2 \times 10^{12}$ |
| *Sample D* | 18 | 3300 | $3.93 \times 10^{12}$ |

TABLE.I. Zhi *et al.*

19